\documentclass[12pt]{iopart}
\usepackage{color}
\usepackage{epstopdf,graphicx,epsfig}
\usepackage{amssymb}
\usepackage{soul} 

\newcommand\foreign[1]{{\it #1\spacefactor=1000}}

\newcommand\ie{\foreign{i.e.}}

\newcommand{\sq}[1]{\left[ {#1} \right]}
\newcommand{\smallfrac}[2]{\mbox{$\frac{#1}{#2}$}}
\newcommand{\half}{\smallfrac{1}{2}}

\newcommand{\expt}[1]{\langle{#1}\rangle}
\newcommand{\dg}{^\dagger}
\newcommand{\D}[1]{{\cal D}\sq{#1}}

\newcommand{\Hc}[1]{{\cal H}\sq{#1}}
\newcommand{\Hcl}[1]{\bar{{\cal H}}\sq{#1}}
\newcommand{\Lc}[1]{{\cal L}_c\sq{#1}}
\newcommand{\Lu}[1]{{{\cal L}}\sq{#1}}
\newcommand{\C}[2]{\left [#1,#2\right ]}

\newcommand{\beq}{\begin{equation}} 
\newcommand{\eeq}{\end{equation}}
\newcommand{\bqa}{\begin{eqnarray}} 
\newcommand{\eqa}{\end{eqnarray}}
\newcommand{\nn}{\nonumber} 

\newcommand{\erf}[1]{Eq.~(\ref{#1})}
\newcommand{\srf}[1]{Sec.~\ref{#1}}
\newcommand{\frf}[1]{Fig.~\ref{#1}}
\renewcommand{\dg}{^\dagger}

\newcommand{\sch}{Schr\"odinger } 
\newcommand{\hei}{Heisenberg }

\newcommand{\xfrac}[2]{{#1}/{#2}}

\renewcommand{\tr}[1]{{\textrm {Tr}}\sq{#1}}

\newcommand{\Lcun}[1]{{\bar{\cal L}}_c\sq{#1}}

\newcommand{\blk}{\color{black}}

\begin{document}

\title{Quantum feedback for rapid state preparation in the presence of control imperfections}

\author{Joshua Combes}
\address{Centre for Quantum Computation and Communication Technology (Australian Research Council),
Centre for Quantum Dynamics, Griffith University, Brisbane, Queensland 4111, Australia}
\address{School of Engineering, Australian National University, Canberra, ACT 0200, Australia}
\address{Center for Quantum Information and Control, University of New Mexico, Albuquerque, NM 87131-0001, USA}
\author{Howard M. Wiseman}
\address{Centre for Quantum Computation and Communication Technology (Australian Research Council),
Centre for Quantum Dynamics, Griffith University, Brisbane, Queensland 4111, Australia}

\begin{abstract}
Quantum feedback control protocols can improve the operation of quantum devices. Here we examine the performance of a purification protocol when there are imperfections in the controls. The ideal feedback protocol produces an $x$ eigenstate from a mixed state in the minimum time, and is known as rapid state preparation. The imperfections we examine include time delays in the feedback loop, finite strength feedback, calibration errors, and inefficient detection. We analyse these imperfections using the Wiseman-Milburn feedback master equation and related formalism. We find that the 
protocol is most sensitive to time delays in the feedback loop. For systems with slow dynamics,  however,  
our analysis suggests that inefficient detection would be the bigger problem. We also show how system imperfections, such as dephasing and damping, can be included in model via the feedback master equation. 
\end{abstract}


\vspace{2pc}
\maketitle

\section{Motivation}
It is well known that imperfections in feedback control protocols--- such as finite strength feedback, inefficient detection, and loop delays --- severely degrade the controllability of some systems \cite{BerWis01, WanWisMil01, ComWisSco10, WisMil10}. In this paper we are interested in these types of control imperfections, as  well as imperfect operations  and non-negligible damping. In the interest of keeping this analysis manageable we restrict ourselves to a single, very simple measurement model, which we now describe.

Measurement-based quantum feedback protocols often consider a system (we will examine a qbit) that has an  interaction with a probe (a current or field) that results in a continuous { quantum non-demolition} (QND) \cite{WisMil10} measurement\footnote{A simplified definition of a QND measurement is one where the measured observable commutes with the time evolution operator of the combined probe field and system, \ie\, $[U,X]=0$.  See Refs.~\cite{WisMil10,BraKha95} for a though treatment. Here we do not consider the control unitary to be included in this $U$ --- its action (after the measurement) certainly does {\em not} commute with $X$, as we will see.} of the system. The change to the system state  $\rho$ during continuous measurement of an observable $X$ is
\begin{equation}\label{SME}
d\rho = dt\Lc{X}\rho \equiv   2\gamma dt \D{X}\rho +\sqrt{2\gamma \eta}dW(t)\Hc{X}\rho.
\end{equation}
The first term, $\D{A} \rho \equiv A\rho A\dg -\half (A\dg A \rho + \rho A\dg A)$, describes the measurement backaction on the state. The final term, $\Hc{A} \rho \equiv  A\rho +\rho A\dg - \tr{(A\dg+ A )\rho}\rho$, represents the refinement to the observer's state of knowledge due to the measurement process \cite{JacSte06,WisMil10}. The positive coefficient $\eta$ is the measurement efficiency. An efficient measurement corresponds to $\eta=1$, while an inefficient measurement corresponds to $\eta<1$. The measurement result in the interval $[t,t+dt)$ is $dR(t) = \sqrt{4\gamma }dt {\expt{X}} + dW(t)/\sqrt{\eta}$, where $dW(t)$ is the Wiener process appearing in \erf{SME} \cite{Jacobs10}. The integrated current is given by $R(t)=\int _{0}^{t} dR(s)$.

In this situation the most general feedback strategy consists of applying a conditional Hamiltonian (which may be a functional of the entire record up to that time) to the system to effect the desired outcome. An important class of feedback strategies are those where the functional of the record is also a function of the best estimate of the system state. These strategies are called Bayesian \cite{WisManWan02} or state-based feedback \cite{DohJac99}. In general, implementing feedback involves engineering driven Hamiltonian terms of the form $\sum_{k} -i dt\alpha_{k}(t)[H_{k},\rho]$ to \erf{SME}. It is common to assume that the controls available are sufficiently strong and fast that one may consider directly controlling the state or measurement basis through a unitary $U$. {(Note that this does not break any rotating wave approximations as we only require $|\alpha|\gg |\gamma|$.)} In this case the conditional state after the measurement and feedback is $\rho(t +dt) = \rho +dt\Lc{\check{X}}\rho $ in the \hei picture with respect to the control unitary, where $\check{X}= U\dg(t)X U(t)$.

Another important class of feedback control strategies can be described by the Wiseman-Milburn Markovian feedback master equation (FBME) \cite{CavMil86, WisMil93,WisMil10}:
\begin{eqnarray}
\nn \dot{\rho} = \Lu \rho &\equiv&  -\frac{i}{\hbar}  \C{\frac{\hbar\sqrt{2\gamma}}{2} (X F+FX)}{\rho} +2\gamma \D{X-\frac{iF}{\sqrt{2\gamma}}}\rho \\ 
&& +\frac{1-\eta}{\eta}\D{F}\rho. \blk \label{MFB_ME_lindblad}
\end{eqnarray}
This can be derived by considering instantaneous feedback of the current via 
the feedback Hamiltonian $H_{\rm fb}(t) = FdR(t)/dt$, and averaging over the measurement noise to obtain 
a deterministic equation \cite{WisMil93,WisMil10}. Here $F$ is a Hermitian operator, heuristically chosen to effect the desired outcome. Markovian feedback is not as powerful as Bayesian feedback; it does however have the advantage that it is experimentally and theoretically simpler.  

In this article we systematically examine the effect of imperfections in quantum feedback protocols with a specific example -- rapid state preparation (which is very closely related to Jacobs' rapid purification of a qbit \cite{Jac0303,Jac0410,WisBou08}). We have chosen this protocol {for a number of reasons}. {Firstly,} rapid state preparation (RSP) has all of the ingredients of more complicated control protocols and experience shows that analysis of RSP often admits analytical solutions. { Secondly RSP, as we shall soon see, is a highly quantum protocol as it requires continually creating (or stabilizing) a state that is a superposition of the {\em pointer basis} \cite{Zur} --- what would, for a larger system, be called a \sch cat state. Consequently studying the effects of imperfections on {\em quantum} control protocols will help delineating in what regimes control will be of practical importance. Finally, much of the excitement about quantum information theory stems from the speed-ups over classical protocols. The RSP  protocol is one quantum control protocol that offers such a speed-up.}

In the ideal limit, the RSP protocol can be achieved with Markovian feedback. To maintain analytic results, in this article we work within the framework of Markovian feedback; if we had worked in the  Bayesian feedback framework we would have been forced to rely on numerics to obtain any results. There is also the possibility of using risk sensitive or robust quantum control \cite{Jam} to examine the problems, but such analysis is also likely to be more complicated than the analysis we present. 

This article is organized as follows. In \srf{SEC:RSP_olc} we present a protocol for state preparation that minimises the use of feedback control to the application of one conditional unitary. In \srf{SEC:RSP_ideal}  we present the general theory for RSP allowing for continuous modulation of a feedback Hamiltonian under ideal conditions.   Then each of these imperfections described above are examined in turn in \srf{SEC:RSP_gen}. Finally we conclude by comparing the relative detrimental effects of each imperfection.

\section{State preparation with one conditional unitary}\label{SEC:RSP_olc}

We begin by presenting the simplest  state preparation protocol for producing a qbit in an $x$ eigenstate, based on continuous monitoring of the $z$ component of angular momentum \ie\ $J_{z}$. {(We choose to use $J_{z}$ rather than the Pauli operators as it will make comparing our results to other results in the literature easy. This is because much of the work in the literature is for qudits.)} It is obvious that this measurement in isolation will only produce eigenstates of $J_{z}$. Thus, to produce an $x$ eigenstate at time $t_{f}$ we assume that it is possible to perform an instantaneous conditional unitary at that time to rotate the Bloch vector to the desired $x$ eigenstate. This is the obvious method for minimising the use of feedback control. We assume here (and in all subsequent protocols) that initially the qbit is in a completely mixed state.

We quantify the effectiveness of this protocol by the probability that it would be verified 
 (by a hypothetical projective measurement) to have produced an $x$-eigenstate at time $t_f$, 
 as motivated in Ref.~\cite{WisBou08}. This probability is linear in the $x$ component of the Bloch 
 vector after the final conditional unitary rotation. This in turn is equal to the length of the Bloch vector 
 prior to that rotation. We take this average length as the figure-of-merit for all of our protocols. 
   
In the present case, it is possible to obtain a closed form expression for the average Bloch vector length as a function of the measurement record by using linear trajectory theory \cite{GoeGra96,Wis9602,JacKni98}. This involves using a linear, and necessarily unnormalized, version of \erf{SME}, called the linear SME 
 \begin{eqnarray}\label{SME_unorm}
 d\bar{\rho}=dt\Lcun{J_{z}}\rho =  2\gamma \, dt\, \D{J_{z}}\rho +\sqrt{2\gamma}\,dR\,\Hcl{J_{z}}\rho,  
 \end{eqnarray}
where  $\Hcl{A} \rho =  A\rho +\rho A\dg$. Here the bar denotes the lack of normalization. Given that $\rho(0)$ and our observable $J_z$ commute, the solution to the SME is  \cite{JacSte06}
\begin{eqnarray}\label{St_rhotildejz}
 \bar{\rho}(R,t)&=&\exp(-4\gamma J_{z}^2t)\exp(2\sqrt{2\gamma}J_{z}R(t))\mathbf{I}/2,
\end{eqnarray}
where $R(t)$ is the integrated photocurrent. Omitting the time dependence on $R$ for compactness, the solution for a qbit can be expressed as
\begin{eqnarray}\label{St_qbitrhotilde}
 \bar{\rho}(R,t)=\frac{e^{-\gamma t}}{2}\left( \begin{array}{c c }
e^{\sqrt{2\gamma}R}      & 0        \\
0      & e^{-\sqrt{2\gamma}R}       \\
\end{array} \right). 
\end{eqnarray}
The norm of \erf{St_rhotildejz}  is $\mathcal{N}=\tr{\bar{\rho} (R, t ) } =e^{-\gamma t} \cosh{(\sqrt{2 \gamma}R)}$. The normalised state is then $\rho(R,t) = \bar{\rho}(R,t)/\mathcal{N}$. The actual probability density for the result $R$ is $\mathcal{P}(R,t)= \tr{\bar{\rho}(R,t)}P(R,t)$. Here 
\beq\label{ostprob}
P(R,t)=  \frac{e^{-R^2/(2t)}}{\sqrt{2\pi t }}
\eeq
has been called the ostensible probability for the result $R$ \cite{Wis9602}. {The ostensible probability is the probability with which the record $R$ should be generated if \erf{SME_unorm} were to be used in place of the normalized SME (\ref{SME}).}

With these results we may now calculate the average Bloch vector length. First we note that the Bloch vector length is given by $2\lambda_{\rm max}[\rho]-1$, where $\lambda_{\rm max}[\rho]$ is the larger eigenvalue of the normalized state $\rho$. From \erf{St_qbitrhotilde}, this will correspond to the probability of being in the $+z$ eigenstate or the $-z$ eigenstate, depending on whether R is positive or negative. Using the symmetry of ${\cal P}(R)$ and \erf{St_qbitrhotilde} the maximum eigenvalue is 
\begin{eqnarray}\label{imp_nfb_qubitt}
 \expt{\lambda_{\rm max}(t)} &=&2\int _{0}^{\infty} \lambda_{\rm max}[\rho(t)]\;\mathcal{N}(R) P(R)dR\nn\\
&=&   \frac{1}{\sqrt{2\pi t}}  \int _{0}^{\infty} e^{-(R-\sqrt{2\gamma}t)^2/2t}dR\nn\\
&=&\half [ 1+\mathrm{erf}(\sqrt{\gamma t})].\label{lambdamax}
\end{eqnarray}
When $t\gg \gamma^{-1}$ (the long-time limit) this expression can be approximated by  
\begin{equation}\label{impqbitlongtime}
\expt{\lambda_{\rm max}(t)}_{\mathrm{LT}} = 1-\frac{1}{2}\frac{ e^{-\gamma t}}{\sqrt{\pi\gamma t }},
\end{equation}
where the subscript LT denotes that this expression is only valid in the long-time limit. We can relate this expression to $\expt{x_{\rm mfb}}$, the mean length of the Bloch vector (or its $x$-component after the final rotation) by
\begin{equation}
\expt{x_{\rm mfb}(t)} = 2\expt{\lambda_{\rm max}(t)}-1.\label{xmfb}
\end{equation}
Here the subscript `mfb' signifies that this protocol minimises the use of feedback control. 
We also refer to this as the open-loop strategy because the  conditional unitary happens after all measurements so it is feedforward.

\section{Ideal rapid state preparation}\label{SEC:RSP_ideal}
Consider again  a qbit in which $J_{z}$ is continuously monitored 
but now we assume that for arbitrarily strong feedback-induced rotation around the $y$ axis is available. This is all that is required for control given that the measured observable picks out the $z$ axis, and that the aim is to maximize $\expt{x(t_f)}$. 
As we will see, the optimal feedback protocol in this situation is of the form described by \erf{MFB_ME_lindblad}, where $H_{\rm fb}(t)dt = F dR(t)$, provided $F(t)$ itself is allowed to be time dependent: $F(t) = \Omega(t) J_y$. Equation (\ref{MFB_ME_lindblad}) will thus provide the basis for the following study.

\label{idealfb}
Given the above senario and allowing for a time dependent feedback strength, $\Omega(t)$, and assuming efficient measurements ($\eta=1$), equation \ref{MFB_ME_lindblad} becomes
\begin{eqnarray}\label{FBMEideal}
\dot{\rho} &=&2\gamma \D{J_{z}-i\frac{\Omega(t)J_{y}}{\sqrt{2\gamma}}}\rho.
\end{eqnarray} 
From \erf{FBMEideal} we may find the the equations of motion for the Bloch components. This is achieved by taking the trace of \erf{FBMEideal} with the component of interest, for example $\dot{x} = \tr{\sigma_{x}\dot{\rho}}$. The resulting equations are
\begin{eqnarray}
\dot{x}&=&-x \gamma  -\frac{ \Omega(t) ^2}{2 }x +\sqrt{2\gamma } \Omega(t)\label{idealx}\\
\dot{y}&=&-\gamma y\label{idealy}\\
\dot{z}&=&-\frac{ \Omega(t) ^2}{2}z\label{idealz}
\end{eqnarray}
By solving $d_{{\Omega}}\dot{x}=0$ for ${\Omega}$, {where $d_{{\Omega}}\equiv d/d\Omega$}, the feedback which locally mazimizes the rate at which the $\dot{x}$ component grows can be found. This gives the optimal feedback strength as a function of $x$: $\Omega_{\rm opt} (x)=\sqrt{2\gamma } /x$. Using this and the initial condition $x(0)=0$ the solution to \erf{idealx} is
\begin{equation}
x_{\rm opt}(t)= \sqrt{1-e^{-2\gamma t}}\label{fbperx}.
\end{equation}
The optimal feedback strength as a function of time is computed by substituting \erf{fbperx} into $\Omega_{\rm opt}$:
\begin{equation}
\Omega_{\rm opt} (t)=\frac{\sqrt{2\gamma }}{ \sqrt{1-e^{-2\gamma t}}}.
 \end{equation}
This is precisely the evolution of Jacobs' rapid purification protocol gives \cite{Jac0303}. Moreover, Wiseman and Bouten have shown that this is the globally optimal control protocol for this problem \cite{WisBou08} (see also Ref.~\cite{BelNegMol09}). To calculate the improvement feedback offers we solve $x_{\rm mfb}(t_{\rm mfb})=x_{\rm opt}(t_{\rm opt})$ for the ratio of $t_{\rm opt}/t_{\rm mfb}$. Then we take the limit that $t_{\rm mfb}\rightarrow \infty$ to give $t_{\rm opt}/t_{\rm mfb} =1/S =1/2$. Thus the asymptotic speed-up (improvement) of the ideal feedback over the minimal control (open-loop) protocol to produce a state with a given fixed value of $x(t_{f})$ very close to one is $S =2$. {Of course, once the desired Bloch vector length is reached we may rotate final state to any desired state, as we have assumed an effectively instantaneous rotation. Thus we may rapidly prepare any qubit state.}

 
\section{Rapid state preparation with imperfections}\label{SEC:RSP_gen}

We now calculate the effect of four different imperfections or limitations on the rapid preparation of an $x$ eignenstate by feedback.

 \subsection{Constant feedback strength}
First we consider the case where the feedback strength is time independent (constant). It is unclear whether such a strategy can rapidly purify at all, and, if it can, whether it affords the same asymptotic advantage as the optimal strategy. In addition this analysis is important experimentally for two reasons. Firstly, a fixed feedback strength is simpler to implement experimentally. Secondly, the optimal feedback strength (15) is unbounded at time zero, so with the analysis here we can probe the usefulness of bounded strength controls.

Based on the asymptotic value of the optimal feedback strength $\Omega_{\rm opt} (\infty)=\sqrt{2\gamma } $, we parameterize the constant feedback strength as $\Omega=\sqrt{2\gamma }\alpha$, where $\alpha\in(-\infty,\infty)$. The $x$ Bloch component becomes 
\begin{eqnarray}\label{constlam}
\nn\dot{x}&=&-x \gamma  -\frac{ {\Omega} ^2}{2 }x +\sqrt{2\gamma } \Omega\\
 &=&-x \gamma  -\gamma \alpha^{2} x +2\gamma\alpha.
\end{eqnarray}
Solving this equation with the initial condition $x(0)=0$ gives
\begin{equation} \label{xlamconstalpha}
x(t)= \frac{2\alpha}{1+\alpha^{2}}(1-e^{-\gamma t(1+\alpha^{2})}).
 \end{equation}
In \frf{fig:constlam}, \erf{xlamconstalpha} is plotted for different values of $\alpha$. Negative values of $\alpha$ result in the $-x$ eigenstate being prepared. When $|\alpha|>1$ the $x$ component of the Bloch vector rapidly increases for $t\ll \gamma^{-1}$ but reaches a steady state value less than one. When $|\alpha|<1$ the Bloch vector increases more slowly, and also asymptotes to a value less than one.
\begin{figure}[h!]
   \centering
\leavevmode \includegraphics[width=0.7\hsize]{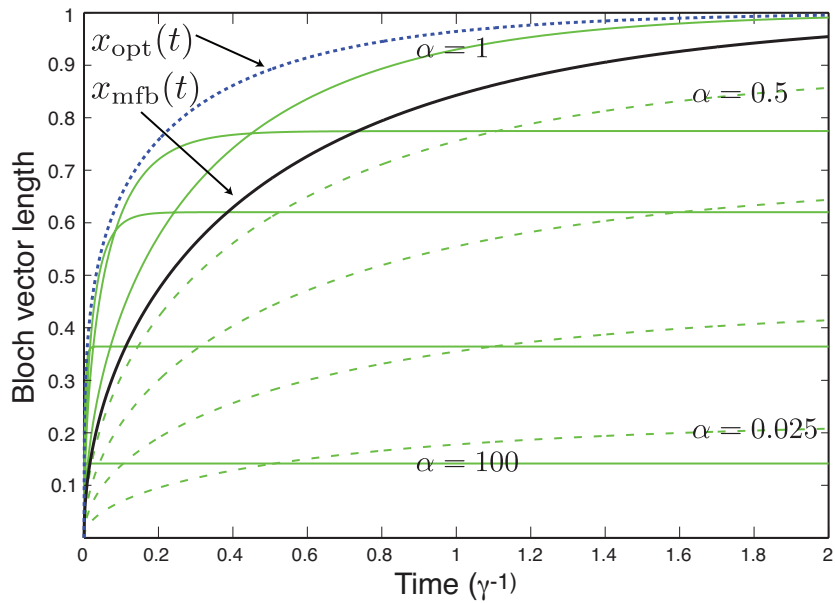}
\caption{The $x$ component of the Bloch vector for different constant feedback strategies. The dotted line (blue) is the ideal feedback i.e. \erf{fbperx}. The thick solid line (black) is for the open-loop control protocol, i.e. \erf{imp_nfb_qubitt}. The remaining lines are for the constant feedback strength strategies. The feedback strength is parameterized by ${\Omega} = \sqrt{2\gamma}\alpha$ where $\alpha\in (-\infty,\infty)$.  From the uppermost curve the solid lines (green) are for $\alpha =1, 3, 5$ and 15. The dashed lines (green), starting from the upper most curve, are for  $\alpha = 0.5, 0.25, 0.1$ and 0.025. 
\label{fig:constlam}
}
\end{figure} 
 
\begin{figure}[h!]
   \centering
\leavevmode \includegraphics[width=0.7\hsize]{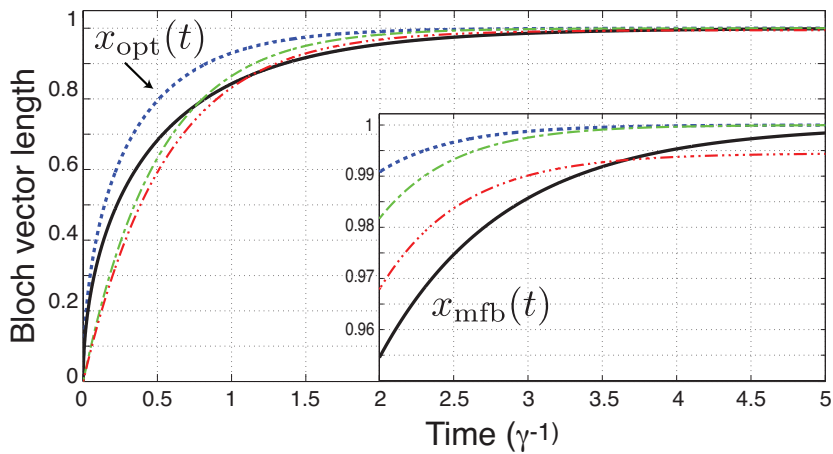}
\caption{A comparison of the time it takes to prepare an $x$ eigenstate for four different strategies. The dotted line (blue) is for the optimal time dependent feedback -- that is, \erf{fbperx}. The thick line (black) is the  Bloch vector length for the open-loop control protocol. The dot-dash (green) line is the asympototicly optimal constant feedback ($\alpha =1$), \erf{xlamconstopt}. The dot-dot-dash (red) line is sub-optimal constant feedback -- that is, \erf{xlamconstalpha} with $\alpha =0.9$. Inset: an exploded version of the same plot.\label{fig:compareconopnfb} }
\end{figure}

The steady state at long-times is $x_{\rm ss}= 2\alpha(1+\alpha^{2})^{-1}$, which is maximized for $\alpha=1$, in which case 
\begin{equation} \label{xlamconstopt}
x(t)= 1-e^{-2\gamma t}.
 \end{equation}
In \frf{fig:constlam} we analyse the short to medium  time regime. \blk The optimal constant feedback begins to out-perform the no feedback case for $t>0.768\gamma^{-1}$. Figure~ \ref{fig:compareconopnfb} shows the long time regime, in which it is appears that \erf{xlamconstopt} approaches the the optimal time dependent feedback curve (\ref{fbperx}). Interestingly, when $\alpha =0.9$ the feedback performs worse than the open-loop control until $t>1.53\gamma^{-1}$. For $1.53\gamma^{-1}<t<3.65\gamma^{-1}$ feedback performs better, and after this interval the feedback again performs worse than the open-loop control. This suggests that an optimal state preparation (or purification) protocol for bounded strength control (or bang-bang control) might switch between periods of feedback and measurement multiple times. 

The only question that remains is whether the optimal constant feedback strength protocol has the same asymptotic advantage as the optimal time dependent feedback. This can be determined by comparing the ratio of the time taken to a fixed Bloch vector length $1-\epsilon$. Taylor-expanding \erf{fbperx} at long-times gives $x_{\rm opt}(t)=1-\half e^{-2\gamma t_{\rm opt}}$. Setting this equal to $1-\epsilon$ and solving for $t_{\rm opt}$ gives the time taken to a fixed Bloch vector length. This process must be repeated for \erf{xlamconstopt}; the time at which \erf{xlamconstopt} reaches $1-\epsilon$ is denoted by $t_{c}$. The ratio of the times is 
\begin{equation} \label{epsspeed}
\frac{t_{c}}{t_{\rm opt}}= \frac{\ln \epsilon}{\ln \epsilon +\ln 2}
 \end{equation}
which approaches 1 as $\epsilon$ approaches zero.
 \subsection{Calibration errors}
Consider now the case in which the applied control has calibration errors. 
 For the optimal time-dependent control a time-dependent calibration error could be modelled by $\Omega (t)=(\sqrt{2\gamma } / \sqrt{1-e^{-2\gamma t}} )(1+\delta)$, where $\delta\in[-1,1]$. Because these errors are systematic it is sensible to assume that $\delta$ is constant. Substituting $\Omega(t)$ into the Bloch equations gives
  \begin{equation}
 \dot{x}=-\gamma x+\sqrt{2\gamma } \Omega(t)(1+\delta )  -\frac{1}{2}  \Omega(t) ^2(1+\delta )^2 x.
\end{equation}
This has the solution 
 \begin{equation}\label{xcalib}
x(t)=\sqrt{\left(1-e^{-2 t \gamma }\right) \left(1-\delta ^2\right)}.
\end{equation}
At long-times this asymptotes to $x_{\rm ss}\sim 1-\delta^{2}/2$ for small $\delta$. In \frf{fig:tdcaler} we plot \erf{xcalib} for different calibration error values. A $25\%$ calibration error performs worse than the open-loop protocol for $t>2.15\gamma^{-1}$. However, provided the calibration error is less than $5\%$ feedback control will out-perform open-loop control for $t\le 5$. For 
comparison we also plot the performance of the asymptotically optimal constant feedback protocol, 
which outperforms the time-dependent feedback with a $5\%$ calibration error  for $t \geq  3.05\gamma^{-1}$. 
 \begin{figure}[h!]
 \centering
\includegraphics[width=0.75\hsize]{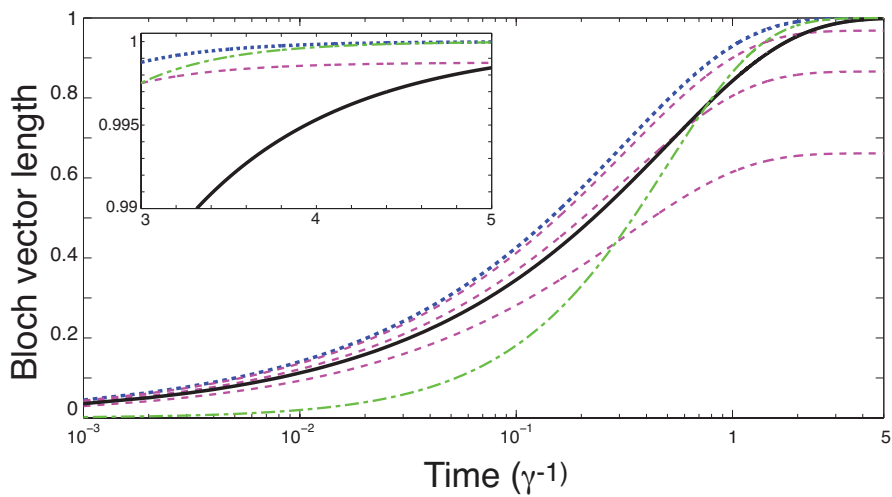}
\caption{The effect of time-dependent calibration errors on RSP. As before the dotted line (blue) is the optimal control; the dot-dashed line (green) is the optimal constant ($\alpha =1$) control; and the thick line (black) is the open-loop control strategy.  The three dashed (magenta) lines from top to bottom are for $\delta = [0.25, 0.5, 0.75]$. The dashed line on the inset figure depicts $\delta = 0.05$. \label{fig:tdcaler} }
\end{figure}

 \subsection{Efficiency and noise on control errors}\label{secineff}
Now consider the case in which the experiment is constrained by a detection inefficiency $\eta$. We analyse this scenario in two ways. First we examine the situation when the experimenter applies the control unaware of detection inefficiencies. The second scenario is more realistic. Here we assume that there is a detection inefficiency and ask what the optimal control in this situation is. Note that detection inefficiency also describes the case when there is random noise on the control \cite{WisMil10}. We note that the effect of detection inefficiency on a related rapid purification protocol was studied in Ref.~\cite{ChiJac08}.

\subsubsection{Oblivious inefficient detection}
When the system is constrained by inefficient detection the equation of motion for the $x$ Bloch component is 
 \begin{equation}\label{dotxineff}
\dot{x}=-x \gamma +\sqrt{2} \sqrt{\gamma } \Omega -\frac{x \Omega ^2}{2 \eta }.
 \end{equation}
If one were oblivious to the detection inefficiency, one would apply the optimal control, $\Omega_{\rm opt}(t)$, from \srf{idealfb}. The solution of the equation of motion under such a control strategy is 
\begin{equation}
x(t)=\sqrt{ \frac{2\eta -1}{\eta}}\sqrt{1-e^{-2\gamma t}}.\label{etaobliv}
 \end{equation}
Clearly the protocol only works for $\eta\in (0.5,1]$ and asymptotically the greatest achievable Bloch vector length  is $x_{\rm ss} = \sqrt{2\eta -1}/\sqrt{\eta}$.

\subsubsection{Optimal control for inefficient detection} 
To optimize the control for inefficient detection one simply takes the derivative of \erf{dotxineff} with respect to ${\Omega}$ and solves to find the optimal control
\begin{equation}
\Omega_{\eta}(t) = \frac{\sqrt{2\gamma }\eta}{x}.
\end{equation}
The solution of the $x$ component using this strategy is
\begin{equation}
x(t)=\sqrt{\eta}\sqrt{\left(1-e^{-2 t \gamma }\right)}.\label{etaopt}
\end{equation}
Asymptotically the steady state value of this stratergy is $x_{\rm ss}=\sqrt{\eta}$. In \frf{fig:etacompare} the steady state values of \erf{etaobliv} and \erf{etaopt}, are plotted for all values of $\eta$. The control optimized for detection inefficiencies, \erf{etaopt}, obviously outperforms the oblivious control case for all $\eta$. 
This is evident in \frf{fig:etaoblivsopt}, where \erf{etaobliv} and \erf{etaopt} are plotted for $\eta=0.85$, although the two strategies are barely distinguishable. With this efficiency, the state preparation becomes worse than that of constant feedback strength, and that of the open-loop strategy, for 
$t \gtrsim \gamma^{-1}$.

  \begin{figure}[h!]
 \centering
\includegraphics[width=0.6\hsize]{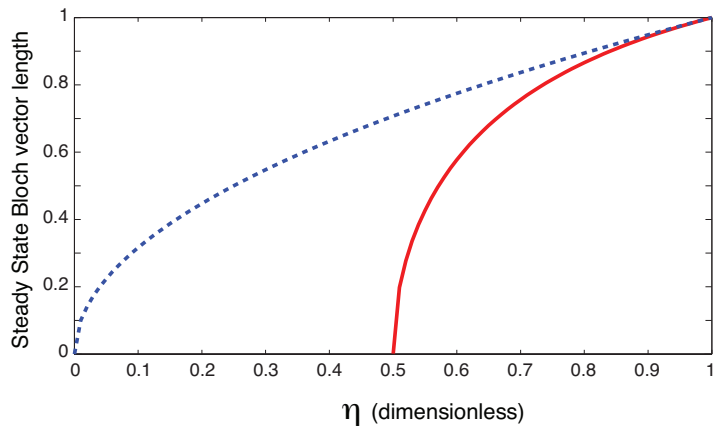}
\caption{A comparison of the steady state inefficiencies for the oblivious control scheme, plotted as the solid line (red), and the optimal control for inefficient detection dotted line (blue).\label{fig:etacompare} }
\end{figure}

   \begin{figure}[h!]
 \centering
\includegraphics[width=0.75\hsize]{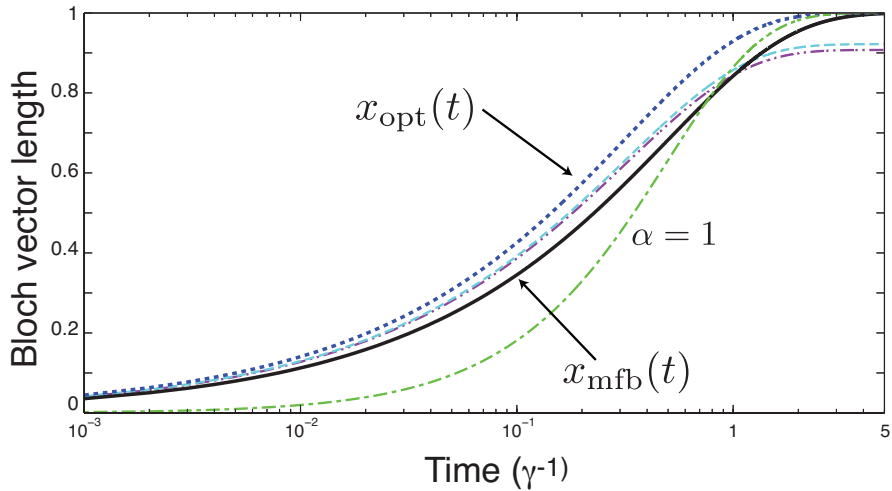}
\caption{A comparison of the dynamics of the  $x$ component of the Bloch vector  at relatively short times for the oblivious control scheme (the lower dot-dot-dashed line) and the optimal control for inefficient detection (the dashed line); $\eta =0.85$. The dotted line is the optimal control; the dot-dashed line is the optimal constant $(\alpha=1)$ control; and the thick line is the open-loop control strategy. \label{fig:etaoblivsopt} }
\end{figure}
 
\subsection{Time delay}

While \erf{MFB_ME_lindblad} is a very general FBME, it assumes instantaneous feedback (that is, zero delay time in the feedback loop). It is possible to include the effects of a small finite time delay in the analysis  while still using a master equation approach \cite{Wis94}. Note however that the resulting equation is not guaranteed to be a valid (Lindblad-form) master equation and will almost certainly give nonphysical results  for $\tau$ too large {with respect to $\gamma^{-1}$}. The equation describing the feedback process including a small time delay $\tau$ is \cite{Wis94}  
\begin{eqnarray}\label{FBMEdelay}
\dot{\rho} &=& \Lu{\rho} -i \tau \sqrt{2\gamma}\C{F}{\Lu{X\rho+\rho X}-X(\Lu \rho)-(\Lu \rho)X}. 
\end{eqnarray} 
Recall  that $\Lu{.}$ was defined in \erf{MFB_ME_lindblad}.
 Substituting $X=J_z$ and $F=\Omega(t)J_{y}$ into \erf{FBMEdelay}, and setting $\eta=1$ gives 
\begin{eqnarray}\label{sdfs}
\dot{\rho} &=&\Lu{\rho} - i  \tau \Omega(t) \sqrt{2\gamma} 
\C{J_{y}}{\Lu{J_{z}\rho+\rho J_{z}}-J_{z}(\Lu \rho)-(\Lu \rho)J_{z}}.
\end{eqnarray} \blk 
From this equation it is straight-forward to calculate the dynamical equations for the Bloch components: 
\begin{eqnarray}
\dot{x}&=&-x \gamma +\sqrt{2\gamma } {\Omega} -\frac{x {\Omega} ^2}{2 }-\frac{\sqrt{2 \gamma } {\Omega} ^3 \tau}{2} \label{fbmext}\\
\dot{y}&=&-\gamma y\label{fbmeyt}\\
\dot{z}&=&-\frac{z {\Omega} ^2}{2  }-2 z {\Omega} ^2 \gamma   \tau\label{fbmezt}.
 \end{eqnarray}
The feedback strength, $\Omega$, must now be chosen in order to rapidly prepare an $x$ eigenstate. 

As for the case of inefficiencies, we choose two scenarios to probe time delays in RSP. The first scenario, presented in \srf{otdelay}, examines the performance of a control protocol which has time delays, but the experimentalist is oblivious to these delays.  The second situation the experimentalist is aware of the feedback delay and compensates for it, which is presented in \srf{aoctdelay}. In both cases we 
take the feedback strength $\Omega$ to be constant, equal to its asymptotically optimal value. 
This is necessary to obtain analytical results, as obtained in all of the other scenarios in this paper.

\subsubsection{Oblivious time delay}\label{otdelay}
Consider an experimentalist who is unaware of time delays in their feedback loop. Due to the cost and difficulties of applying time dependent controls the experimentalist decides to apply asymptotically optimal time independent control $\Omega = \sqrt{2\gamma}$ in their RSP procedure. The solution of \erf{fbmext} in this scenario is 
\begin{equation}\label{xtimedelay}
x(t)=\left(1-e^{-2 t \gamma }\right) \left(1-\gamma  \tau \right).
\end{equation}
The steady state solution is  $x_{\rm ss}= 1-\gamma  \tau $. In \frf{fig:tdelay} \erf{xtimedelay} is plotted for different time delays. %
\begin{figure}[!h]
 \centering
\includegraphics[width=0.7\hsize]{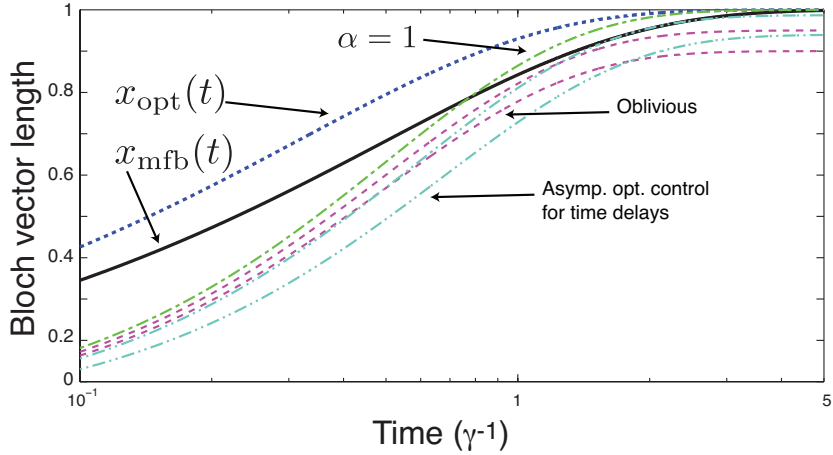}
\caption{A comparison of different feedback strategies for different time delays. The dotted line is the optimal control; the dot-dash line is the optimal constant $(\alpha=1)$ control; and the think line is the measurement alone strategy. The dashed lines (magenta) from top to bottom are for feedback with oblivious time delay with $\tau = 5\times10^{-2}\gamma^{-1},10^{-1}\gamma^{-1}$. The dot-dot-dashed lines (cyan) from top to bottom are for the asymptotically optimal control with time delays $\tau = 5\times10^{-2}\gamma^{-1},10^{-1}\gamma^{-1}$.\label{fig:tdelay} }
\end{figure}
 
\subsubsection{Asymptotically optimal control for time delays}\label{aoctdelay} 

Consider first a local-in-time optimization of \erf{fbmext}. This \blk is again achieved by solving $d_{{\Omega}} \dot{x}=0$ for ${\Omega}$, which gives 
$\Omega_{\rm opt}^{\pm} ={-x\pm\sqrt{x^2+12  \gamma  \eta \tau }}/({3 \sqrt{2\gamma}   \tau })$.
A second derivative test shows that $\Omega_{\rm opt}^{+}$ is the local maximum, thus we define $\Omega_{\rm opt}\equiv \Omega_{\rm opt}^{+} $. As \erf{FBMEdelay} is a perturtabative correction to order $\tau$, it is only sensible to consider the terms to order $\tau$: 
\begin{equation}
\Omega_{\rm opt} =\frac{\sqrt{2\gamma }\eta}{x}-\frac{3 \sqrt{2\gamma}\eta^{2}  \gamma  \tau}{x^3}+O(\tau^{2}).\label{fboptst}
\end{equation}

If we restrict, as in the oblivious case, to constant feedback,\footnote{The alternative would be to solve the problem numerically. The relevant equation with locally optimal feedback (\ref{fboptst}) is $\dot{x}=-x \gamma +\xfrac{\gamma  \eta }{x}-\xfrac{2 \gamma ^{2} \eta ^2 \tau }{x^3}$.} then optimal asymptotic value in the presence of time delays is thus $\Omega =\sqrt{2\gamma}(1-3\gamma \tau)$. Under this control statergy the solution of $\dot{x}$ is
\begin{equation}\label{timedelayz}
x(t)=\frac{2(1-3 \gamma  \tau )}{2(1-3 \gamma  \tau) +9 \gamma ^2 \tau ^2}\left(1-e^{-2 \gamma t +6  \gamma ^2 \tau t -9  \gamma ^3 \tau^2 t}\right)
\end{equation}
It is apparent from \frf{fig:tdelay} that the asymptotically optimal control significantly outperforms the oblivious senerio for $t\gtrsim \gamma^{-1}$. In fact it gives a final Bloch vector length of almost unity. This is because the coefficient \erf{timedelayz}, to first order in $\tau$, is unity. However the terms in the exponent slow the purification, so that no improvement over the open-loop case 
is found at any time even for $\tau = 0.05 \gamma^{-1}$.

\section{Comparison of imperfections}

In order to compare the imperfections we determine the allowed parameter ranges for the different imperfections that will enable the feedback control case to perform better than the minimal control (open-loop) protocol at $t=2\gamma^{-1}$ and $t=4\gamma^{-1}$. Although the choice of times is arbitrary, the first time chosen is, roughly, when the long-time limit ($t \gg \gamma^{-1}$) approximation begins to be reasonable, while the second is well in the asymptotic regime. Table~\ref{impercontrol} summarizes the results.

\begin{table}[h!]
\begin{center}\begin{tabular}{| l | l | l |}
\hline\hline
\textbf{Control Imperfection}  &  \textbf{Parameter Range A}  & \textbf{Parameter Range B} \\
\hline Constant FB strength                               & $0.956\le \alpha \le 1.189$                   &  $0.9415\le \alpha \le 1.0678$  \\
\hline Time dep. cal. errors                                & $\phantom{.887}0\le \delta \le 0.165$     &  $\phantom{.8877}0\le \delta \le \, 0.0659$ \\
\hline Inefficient det. oblivious FB                       & $0.973\le \eta \le 1$                             &    $0.9956\le \eta \le 1$\\
\hline Inefficient det. optimal FB                          & $ 0.972\le \eta \le 1$                           & $0.9956\le \eta \le 1$\\
\hline Time delay $(\gamma^{-1})$ oblivious FB   & $ \phantom{.887}0\le \tau \le 0.0045$     & $ \phantom{.8877}0\le \tau \le 0.0020$\\ 
\hline Time delay $(\gamma^{-1})$  optimal FB    & $ \phantom{.887}0\le \tau \le 0.0146$      &$ \phantom{.8877}0\le \tau \le 0.0195$\\
\hline\hline
\end{tabular} \caption{The parameter ranges for which imperfect feedback performs better than the minimal control protocol at a fixed time. The first column describes the type of control imperfection. The second column, Parameter Range A, gives the approximate values for the parameter (imperfection) under consideration at $t=2\gamma^{-1}$. The third column, Parameter Range B, gives the approximate values for the parameter (imperfection) under consideration at $t=4\gamma^{-1}$.\label{impercontrol}}
\end{center}
\end{table}

\begin{table}[!h]
\begin{center}\begin{tabular}{| l | l | l |}
\hline\hline
\textbf{Control Imperfection}  &  \textbf{Parameter Range A}  & \textbf{Parameter Range B} \\
\hline Constant FB strength                               & $0.922\le \alpha \le 1.135$                   &  $0.9860\le \alpha \le 1.0141$  \\
\hline Time dep. cal. errors                                & $\phantom{.887}0\le \delta \le 0.1246$     &  $\phantom{.8877}0\le \delta \le \, 0.01412$ \\
\hline Inefficient det. oblivious FB                       & $0.9846\le \eta \le 1$                             &    $0.9998\le \eta \le 1$\\
\hline Inefficient det. optimal FB                          & $0.9844\le \eta \le 1$                           & $0.9998\le \eta \le 1$\\
\hline Time delay $(\gamma^{-1})$ oblivious FB   & $ \phantom{.887}0\le \tau \le 0.00547$     & $ \phantom{.8877}0\le \tau \le 0.0000996$\\ 
\hline Time delay $(\gamma^{-1})$  optimal FB    & $ \phantom{.887}0\le \tau \le 0.02598$      &$ \phantom{.8877}0\le \tau \le 0.004612$\\
\hline\hline
\end{tabular} \caption{The parameter ranges for which imperfect feedback performs better than the minimal control protocol at a  fixed Bloch vector length. The first column describes the type of control imperfection. The second column, Parameter Range A, gives the approximate values for the parameter (imperfection) under consideration at $x=1-10^{-2}$. The third column, Parameter Range B, gives the approximate values for the parameter (imperfection) under consideration at $x=1-10^{-4}$.\label{impercontrol2}}
\end{center}
\end{table}

From \frf{fig:constlam} it was clear that $\alpha \gg 1$ is required for short time evolution to perform like the open-loop evolution, but the dynamics induced by a large $\alpha$ caused the steady state Bloch vector length to be poor. Thus the second row in Table \ref{impercontrol} suggests that strong feedback is not necessary as $\alpha$ is of order 1. This implies that bounded strength control protocols can work in RSP. The third row in Table (\ref{impercontrol}) suggests that the feedback protocol is not very sensitive to calibration errors, as a $16.5\%$ error in the calibration seems rather high. 

The fourth and fifth rows of Table \ref{impercontrol} show the effect of detection inefficiencies on RSP. Even in quantum optics with high-efficiency homodyne detectors, the overall efficiency is seldom greater than $\eta <0.9$ (see e.g. Ref.~\cite{Whe10}).   
Thus, unfortunately, the RSP protocol is likely to be severely compromised by detection inefficiencies, regardless of whether one is aware of them or not. These rows also represent the effects of white noise on the control.

 The final row indicates that RSP is most sensitive to time delays. Outside the small $\tau$ regime it is unlikely that this sort of RSP will be feasible using Markovian or Bayesian feedback.\footnote{We note, however, that the open-loop scheme presented in Ref. \cite{ComWisSco10} allows sub-optimal RSP with a single conditional unitary at the end, as in the minimal control protocol of Sec.~\ref{SEC:RSP_olc}.} If, however, a particular physical system has $\gamma^{-1}$ which is of the order of a Meghertz (or smaller) then feedback delays due to classical circuitry could be made much smaller than $\gamma^{-1}$. In this case RSP is remarkably robust to time delays, in the sense that the asymptotic $x_{\rm ss}$ differs from unity only in second order in the delay $\tau$.

In Table \ref{impercontrol2} we consider the parameter ranges that allow feedback to perform better than the minimal control protocol at a fixed Bloch vector length. We consider two lengths:  $x=1-10^{-2}$ (the second column) and $x=1-10^{-4}$ (the third column). The same trends described above also hold for this table.

\section{Discussion}
The most important conclusion of our analysis is that 
quantum feedback control for rapid state preparation is quite sensitive to detection inefficiencies. In particular, locally optimizing the feedback, taking into account the efficiency $\eta$, limits the 
length of the Bloch vector in the long-time limit to $\sqrt{\eta}$.
However, our analysis also turned up some encouraging conclusions. Surprisingly, control in an unbiased basis seems robust to delays in the feedback loop, at least in the sense that 
the asymptotic Bloch differs from unity only in second order in the delay $\tau$.

The methodology presented in this article, approximating Bayesian feedback with Markovian feedback, is not restricted to studying imperfections in the control. It also allows one to study the performance of control protocols with system imperfections. For example, {consider performing RSP when the } system has isotropic dephasing noise at rate $\Gamma_{\rm iso}$ and the decay rate from the excited state is $\Gamma_{\rm d}$. {The FBME in this situation is: 
\begin{equation}
 \dot{\rho}= \Lu{\rho}+{\sum_{i\in x,y,z} \Gamma_{\rm iso}\D{\sigma_{i}}\rho+  \Gamma_{\rm d}}\D{\sigma_{-}}\rho. 
 \end{equation}
 From this equation it is easy to derive an equation for $x(t)$}
\begin{equation}
 x(t)=\sqrt{\frac{2\gamma  \eta }{2 \gamma +8  \Gamma_{\rm iso}+\Gamma_{\rm d} }}\sqrt{1-e^{(-2 t \gamma -8 t  \Gamma_{\rm iso} -\Gamma_{\rm d} t)}}. 
\end{equation}
Here we are modelling locally optimal rapid state preparation, with detection inefficiency included as a control imperfection. The main point is that, by approximating Bayesian feedback with Markovian feedback we can begin to analyse {\em analytically} the performance of feedback protocols with many different control or system imperfections. We note that the effect of $x$ dephasing and decay from the excited state was studied in Ref.~\cite{LiJac09}.
The advantage of our method is two fold: firstly our analysis is algebraically simpler; and  secondly we derive time-dependent solutions, not just steady state values.

In future work we think it would also be interesting to see if the Markovian approach to rapid $x$-eigenstate preparation presented here could  be generalized to  all qbit states (by analogy with the stabilization of a qbit under monitored spontaneous emission in Refs.~\cite{HofMahHes98,WisManWan02}) with and without imperfections. 

{\em Acknowledgements:} This research was conducted by the Australian Research Council Centre of Excellence for Quantum Computation and Communication Technology (project number CE110001029).  JC also acknowledges support from National Science Foundation Grant No. PHY-0903953 and Office of Naval Research Grant No. N00014-11-1-008.

\end{document}